\documentclass{revtex4}
\usepackage{amsfonts}
\usepackage{amsmath}

\input{tcilatex}

\begin{document}

\title[ ]{}
\title{Higher dimensional thin-shell wormholes in
Einstein-Yang-Mills-Gauss-Bonnet gravity}
\author{S. Habib Mazharimousavi}
\email{habib.mazhari@emu.edu.tr}
\author{M. Halilsoy}
\email{mustafa.halilsoy@emu.edu.tr}
\author{Z. Amirabi}
\email{zahra.amirabi@emu.edu.tr}
\affiliation{Department of Physics, Eastern Mediterranean University, G. Magusa, north
Cyprus, Mersin 10, Turkey. }

\begin{abstract}
We present thin-shell wormhole solutions in Einstein-Yang-Mills-Gauss-Bonnet
(EYMGB) theory in higher dimensions $d\geq 5$. Exact black hole solutions
are employed for this purpose where the radius of thin-shell lies outside
the event horizon. For some reasons the cases $d=5$ and $d>5$ are treated
separately. The surface energy-momentum of the thin-shell creates surface
pressures to resist against collapse and rendering stable wormholes
possible. We test the stability of the wormholes against spherical
perturbations through a linear energy-pressure relation and plot stability
regions. Apart from this restricted stability we investigate the possibility
of normal (i.e. non-exotic) matter which satisfies the energy conditions.
For negative values of the Gauss-Bonnet (GB) parameter we obtain such
physical wormholes.

\emph{Dedicated to the memory of Rev. Ibrahim EKEN (1927-2010) of Turkey}.
\end{abstract}

\maketitle

\section{INTRODUCTION}

One of the challenging problems in general relativity is to construct
viable, traversable wormholes \cite{1,2} from curvature of spacetime and
physically meaningful energy-momenta. Most of the sources to support
wormholes to date, unfortunately consists of exotic matter which violates
the energy conditions \cite{3}. More recently, however, there are examples
of thin-shell wormholes that resist against collapse when sourced entirely
by physical (normal) matter satisfying the energy conditions \cite{4}. From
this token, it has been observed that pure Einstein's gravity consisting of
Einstein-Hillbert (EH) action with familiar sources alone doesn't suffice to
satisfy the criteria required for normal matter. This leads automatically to
taking into account the higher curvature corrections known as the Lovelock
hierarchy \cite{5}. Most prominent term among such higher order corrections
is the Gauss-Bonnet (GB) term to modify the EH Lagrangian. There is already
a growing literature on Einstein-Gauss-Bonnet (EGB) gravity and wormhole
constructions in such a theory.

In this paper we intend to fill a gap in this line of thought which concerns
Einstein-Yang-Mills (EYM) theory amended with the GB term. More
specifically, we wish to construct thin-shell wormholes that are supported
by normal (i.e. non-exotic) matter. To this end, we first construct higher
dimensional ($d\geq 5$) exact black hole solutions in EYMGB theory. This we
do by employing the higher dimensional Wu-Yang ansatz which has been
described elsewhere \cite{6,7}. The distinctive point with this particular
ansatz is that the YM invariant emerges with the same power, irrespective of
the spacetime dimensionality. In this regard EYM solution becomes simpler in
comparison with the Einstein-Maxwell (EM) solutions. This motivates us to
seek for thin-shell wormholes by cutting / pasting method in EYM theory.

Another point of utmost importance is the GB parameter ($\alpha $), whose
sign plays a crucial role in the positivity of energy of the system.
Although in string theory this parameter is chosen positive for some valid
reasons, when it comes to the subject of wormholes our choice favors the
negative values ($\alpha <0$), for the GB parameter. One more item that we
consider in detail in this study is to investigate the stability of such
wormholes against linear perturbations when the pressure and energy density
are linearly related.

The exact solution to EYMGB gravity that we shall employ in this paper were
established before \cite{6,7}. Our line element is chosen in the form \cite%
{7} 
\begin{equation}
ds^{2}=-f\left( r\right) dt^{2}+\frac{dr^{2}}{f\left( r\right) }%
+r^{2}d\Omega _{d-2}^{2},
\end{equation}%
in which $f\left( r\right) $ is the only metric function and 
\begin{equation}
d\Omega _{d-2}^{2}=d\theta _{1}^{2}+\underset{i=2}{\overset{d-2}{\tsum }}%
\underset{j=1}{\overset{i-1}{\tprod }}\sin ^{2}\theta _{j}\;d\theta _{i}^{2},
\end{equation}%
where%
\begin{equation*}
0\leq \theta _{d-2}\leq 2\pi ,0\leq \theta _{i}\leq \pi ,\text{ \ \ }1\leq
i\leq d-3.
\end{equation*}%
According to the higher dimensional Wu-Yang ansatz the YM potential is
chosen as%
\begin{align}
\mathbf{A}^{(a)}& =\frac{Q}{r^{2}}C_{\left( i\right) \left( j\right)
}^{\left( a\right) }\ x^{i}dx^{j},\text{ \ \ }Q=\text{YM magnetic charge, \ }%
r^{2}=\overset{d-1}{\underset{i=1}{\sum }}x_{i}^{2}, \\
2& \leq j+1\leq i\leq d-1,\text{ \ and \ }1\leq a\leq \left( d-2\right)
\left( d-1\right) /2,  \notag \\
x_{1}& =r\cos \theta _{d-3}\sin \theta _{d-4}...\sin \theta _{1},\text{ }%
x_{2}=r\sin \theta _{d-3}\sin \theta _{d-4}...\sin \theta _{1},  \notag \\
\text{ }x_{3}& =r\cos \theta _{d-4}\sin \theta _{d-5}...\sin \theta _{1},%
\text{ }x_{4}=r\sin \theta _{d-4}\sin \theta _{d-5}...\sin \theta _{1}, 
\notag \\
& ...  \notag \\
x_{d-2}& =r\cos \theta _{1},  \notag
\end{align}%
where $C_{\left( b\right) \left( c\right) }^{\left( a\right) }$ is the
non-zero structure constants \cite{8}. By this choice the YM invariant $%
\mathcal{F}$ reduces to a simple form%
\begin{equation}
\mathcal{F}=\mathbf{Tr}(F_{\lambda \sigma }^{\left( a\right) }F^{\left(
a\right) \lambda \sigma })=\frac{\left( d-3\right) }{r^{4}}Q^{2},
\end{equation}%
which yields the energy-momentum tensor%
\begin{equation}
T_{\text{ }\mu }^{\nu }=-\frac{1}{2}\mathcal{F}\text{diag}\left[ 1,1,\kappa
,\kappa ,..,\kappa \right] ,\text{ \ and \ }\kappa =\frac{d-6}{d-2}.
\end{equation}%
Accordingly, the field equations are (without a cosmological term) 
\begin{equation}
G_{\mu \nu }^{E}+\alpha G_{\mu \nu }^{GB}=T_{\mu \nu },
\end{equation}%
where 
\begin{equation}
G_{\mu \nu }^{GB}=2\left( -R_{\mu \sigma \kappa \tau }R_{\quad \nu }^{\kappa
\tau \sigma }-2R_{\mu \rho \nu \sigma }R^{\rho \sigma }-2R_{\mu \sigma }R_{\
\nu }^{\sigma }+RR_{\mu \nu }\right) -\frac{1}{2}\mathcal{L}_{GB}g_{\mu \nu }%
\text{ ,}
\end{equation}%
$\alpha $ is the GB parameter and GB Lagrangian $\mathcal{L}_{GB}$ is given
by 
\begin{equation*}
\mathcal{L}_{GB}=R_{\mu \nu \gamma \delta }R^{\mu \nu \gamma \delta
}-4R_{\mu \nu }R^{\mu \nu }+R^{2}.
\end{equation*}

The exact solutions which we shall use throughout this paper are \cite{6,7}%
\bigskip 
\begin{equation}
f_{\pm }\left( r\right) =\left\{ 
\begin{array}{cc}
1+\frac{r^{2}}{4\alpha }\left( 1\pm \sqrt{1+\frac{32\alpha M_{ADM}}{3r^{4}}+%
\frac{16\alpha Q^{2}\ln r}{r^{4}}}\right) ,\text{ \ \ \ } & d=5 \\ 
1+\frac{r^{2}}{2\tilde{\alpha}}\left( 1\pm \sqrt{1+\frac{16\tilde{\alpha}%
M_{ADM}}{r^{d-1}\left( d-2\right) }+\frac{4\left( d-3\right) \tilde{\alpha}%
Q^{2}}{\left( d-5\right) r^{4}}}\right) ,\text{ \ \ \ } & d\geq 6%
\end{array}%
\right. ,\text{ \ \ \ \ \ \ \ \ \ \ \ \ \ }
\end{equation}%
in which $\tilde{\alpha}=\left( d-3\right) \left( d-4\right) \alpha ,$ with
the GB parameter $\alpha $. Here $M_{ADM}$ stands for the usual ADM mass of
the black hole and $Q$ is the YM charge. When compared with Ref.s \cite{6}
(for $d=5$) and \cite{7} (for $d>5$) the meaning of $M_{ADM}$ implies that $%
M_{ADM}=\frac{3}{2}\left( m+2\alpha \right) $ and $M_{ADM}=\frac{1}{4}%
m\left( d-2\right) ,$ respectively. Let us also add that in Ref. \cite{7} we
set $Q=1$ through scaling. The crucial point in our solution is that the YM
term under the square root has a fixed power $\frac{1}{r^{4}}$ for all d$%
\geq 6.$\ As it can be checked, the negative branch gives the correct limit
of higher dimensional black hole solution in EYM theory of gravity if $%
\alpha \rightarrow 0,$ and therefore in the sequel we only consider this
specific case. We also notice that for negative $\alpha $ there exists a
curvature singularity at $r=r_{\circ }$ where $r_{\circ }$ is the smallest
radius of which, for $r>r_{\circ },$ inside the square root is positive. For 
$\alpha >0,$ although for $d\geq 6$ there is no curvature singularity, for $%
d=5$ it depends on the value of the free parameters (i.e., $\tilde{\alpha},$ 
$Q,$ $M_{ADM}$) to result in a curvature singularity.

Here, in order to explore the physical properties of the above solutions we
investigate some essential thermodynamic quantities. Since $d=5$ case has
been studied elsewhere \cite{9} we shall concentrate on $d\geq 6.$

Radius of the event horizon (i.e., $r_{h}$) of the negative branch black
hole $f_{-}\left( r\right) ,$ with positive $\alpha $ is the maximum root of 
$f_{-}\left( r_{h}\right) =0.$ It is not difficult to show that in terms of
event horizon radius one can write%
\begin{equation}
M_{ADM}=\frac{\left( d-2\right) }{4}\left[ \left( \tilde{\alpha}%
+r_{h}^{2}\right) -\frac{\left( d-3\right) }{\left( d-5\right) }Q^{2}\right]
r_{h}^{d-5}.
\end{equation}%
Also we find the Hawking temperature $T_{H}$ in terms of $r_{h},$ i.e., 
\begin{equation}
T_{H}=\frac{1}{4\pi }f^{\prime }\left( r_{h}\right) =\frac{\left( d-3\right)
\left( r_{h}^{2}-Q^{2}\right) +\tilde{\alpha}\left( d-5\right) }{4\pi
r_{h}\left( 2\tilde{\alpha}+r_{h}^{2}\right) }.
\end{equation}%
To complete our thermodynamical quantities we use the standard definition of
the specific heat capacity with the constant charge 
\begin{equation}
C_{Q}=T_{H}\left( \frac{\partial S}{\partial T_{H}}\right) _{Q},
\end{equation}%
in which $S$ is the standard entropy defined as 
\begin{equation}
S=\frac{A}{4}=\frac{\left( d-1\right) \pi ^{\frac{d-1}{2}}}{4\Gamma \left( 
\frac{d+1}{2}\right) }r_{h}^{d-2},
\end{equation}%
to show the possible thermodynamical phase transition. After some
manipulation we find 
\begin{equation}
\begin{tabular}{l}
$C_{Q}=\frac{\left( d-2\right) \left( d-1\right) \left( 2\tilde{\alpha}%
+r_{h}^{2}\right) \pi ^{\frac{d-1}{2}}r_{h}^{d-2}\left[ \left( d-5\right) 
\tilde{\alpha}+\left( d-3\right) \left( r_{h}^{2}-Q^{2}\right) \right] }{%
4\Gamma \left( \frac{d+1}{2}\right) \left\{ 2\tilde{\alpha}\left[
Q^{2}\left( d-3\right) -\tilde{\alpha}\left( d-5\right) \right] +\left[
3Q^{2}\left( d-3\right) -\tilde{\alpha}\left( d-9\right) \right]
r_{h}^{2}-\left( d-3\right) r_{h}^{4}\right\} }.$%
\end{tabular}%
\end{equation}%
The phase transition is taking place at the real and positive root(s) of the
denominator, i.e., 
\begin{equation}
\begin{tabular}{l}
$2\tilde{\alpha}\left[ Q^{2}\left( d-3\right) -\tilde{\alpha}\left(
d-5\right) \right] +\left[ 3Q^{2}\left( d-3\right) -\tilde{\alpha}\left(
d-9\right) \right] r_{h}^{2}-\left( d-3\right) r_{h}^{4}=0.$%
\end{tabular}%
\end{equation}%
One can show that under the condition%
\begin{equation}
\frac{Q^{2}}{\tilde{\alpha}}<\frac{7d-39}{9\left( d-3\right) }
\end{equation}%
there is no phase transition, while if 
\begin{equation}
\frac{7d-39}{9\left( d-3\right) }<\frac{Q^{2}}{\tilde{\alpha}}<\frac{d-5}{d-3%
}
\end{equation}%
we will observe two phase transitions. Finally upon choosing 
\begin{equation}
\frac{d-5}{d-3}\leq \frac{Q^{2}}{\tilde{\alpha}}
\end{equation}%
there exists only one phase transition. Also, if $\frac{Q^{2}}{\tilde{\alpha}%
}=\frac{7d-39}{9\left( d-3\right) }$ one phase transition occurs at $r_{h}=%
\sqrt{\frac{6\left( d-3\right) }{7d-39}Q^{2}}$ . These results show that the
dimensionality of spacetime plays a crucial role in the thermodynamical
behavior of the EYMGB system.

For negative $\alpha $ in the negative branch we write $\tilde{\alpha}%
=-\left\vert \tilde{\alpha}\right\vert $ and therefore the horizon radius $%
r_{h}$ is given by solving 
\begin{equation}
1-\frac{2\left\vert \tilde{\alpha}\right\vert }{r_{h}^{2}}=\sqrt{1-\frac{%
16\left\vert \tilde{\alpha}\right\vert M_{ADM}}{r_{h}^{d-1}\left( d-2\right) 
}-\frac{4\left( d-3\right) \left\vert \tilde{\alpha}\right\vert Q^{2}}{%
\left( d-5\right) r_{h}^{4}}}.
\end{equation}

The method of establishing the thin-shell wormhole, based on the black hole
solutions given in (8), follows the standard procedure which has been
employed in many recent works \cite{4}.

\section{Dynamic thin-shell wormholes in $d-$dimensions}

The method of establishing a thin-shell wormhole in the foregoing geometry
goes as follows. We cut two copies of the EYMGB spacetime 
\begin{equation}
M^{\pm }=\left\{ r^{_{\pm }}\geq a,\text{ \ }a>r_{h}\right\}
\end{equation}%
and paste them at the boundary hypersurface $\Sigma ^{\pm }=\left\{ r^{_{\pm
}}=a,\text{\ }a>r_{h}\right\} $. These surfaces are identified on $r=a$ with
a surface energy-momentum of a thin-shell whose radius coincides also with
the throat radius such that geodesic completeness holds for $M=M^{+}\cup
M^{-}$. Following the Darmois-Israel formalism \cite{10} in terms of the
original coordinates $x^{\gamma }=\left( t,r,\theta _{1},\theta
_{2},...\right) $ (i.e. in $M$) the induced metric $\xi ^{i}=\left( \tau
,\theta _{1},\theta _{2},...\right) ,$ on $\Sigma $ is given by (Latin
indices run over the induced coordinates i.e., $\left\{ 1,2,3,..,d-1\right\} 
$ and Greek indices run over the original manifold's coordinates i.e., $%
\left\{ 1,2,3,..,d\right\} $)%
\begin{equation}
g_{ij}=\frac{\partial x^{\alpha }}{\partial \xi ^{i}}\frac{\partial x^{\beta
}}{\partial \xi ^{j}}g_{\alpha \beta }.
\end{equation}%
Here $\tau $ is the proper time and 
\begin{equation}
g_{ij}=\text{diag}\left( -1,a^{2},a^{2}\sin ^{2}\theta ,a^{2}\sin ^{2}\theta
\sin ^{2}\phi ,...\right) ,
\end{equation}%
while the extrinsic curvature is defined by 
\begin{equation}
K_{ij}^{\pm }=-n_{\gamma }^{\pm }\left( \frac{\partial ^{2}x^{\gamma }}{%
\partial \xi ^{i}\partial \xi ^{j}}+\Gamma _{\alpha \beta }^{\gamma }\frac{%
\partial x^{\alpha }}{\partial \xi ^{i}}\frac{\partial x^{\beta }}{\partial
\xi ^{j}}\right) _{r=a}.
\end{equation}%
It is assumed that $\Sigma $ is non-null, whose unit $d-$normal in $M^{\pm }$
is given by 
\begin{equation}
n_{\gamma }=\left( \pm \left\vert g^{\alpha \beta }\frac{\partial F}{%
\partial x^{\alpha }}\frac{\partial F}{\partial x^{\beta }}\right\vert
^{-1/2}\frac{\partial F}{\partial x^{\gamma }}\right) _{r=a},
\end{equation}%
in which $F$ is the equation of the hypersurface $\Sigma ,$ i.e. 
\begin{equation}
\Sigma :F\left( r\right) =r-a\left( \tau \right) =0.
\end{equation}

The generalized Darmois-Israel conditions on $\Sigma $ determines the
surface energy-momentum tensor $S_{ab}$ which is expressed by \cite{11}%
\begin{equation}
S_{i}^{j}=-\frac{1}{8\pi }\left( \left\langle K_{i}^{j}\right\rangle
-K\delta _{i}^{j}\right) -\frac{\alpha }{16\pi }\left\langle
3J_{i}^{j}-J\delta _{i}^{j}+2P_{imn}^{\;\;\;\;j}K^{mn}\right\rangle .
\end{equation}%
Here a bracket implies a jump across $\Sigma .$ The divergence-free part of
the Riemann tensor $P_{abcd}$ and the tensor $J_{ab}$ (with trace $%
J=J_{a}^{a}$) are given by 
\begin{align}
P_{imnj}& =R_{imnj}+\left( R_{mn}g_{ij}-R_{mj}g_{in}\right) -\left(
R_{in}g_{mj}-R_{ij}g_{mn}\right) +\frac{1}{2}R\left(
g_{in}g_{mj}-g_{ij}g_{mn}\right) , \\
J_{ij}& =\frac{1}{3}\left[
2KK_{im}K_{j}^{m}+K_{mn}K^{mn}K_{ij}-2K_{im}K^{mn}K_{nj}-K^{2}K_{ij}\right] .
\end{align}%
By employing these expressions through (25) we find the energy density and
surface pressures for a generic metric function $f\left( r\right) ,$ with $%
r=a\left( \tau \right) .$ The results are given by 
\begin{gather}
\sigma =-S_{\tau }^{\tau }=-\frac{\Delta \left( d-2\right) }{8\pi }\left[ 
\frac{2}{a}-\frac{4\tilde{\alpha}}{3a^{3}}\left( \Delta ^{2}-3\left( 1+\dot{a%
}^{2}\right) \right) \right] , \\
\begin{tabular}{l}
$S_{\theta _{i}}^{\theta _{i}}=p=$ \\ 
$\frac{1}{8\pi }\left\{ \frac{2\left( d-3\right) \Delta }{a}+\frac{2\ell }{%
\Delta }-\frac{4\tilde{\alpha}}{3a^{2}}\left[ 3\ell \Delta -\frac{3\ell }{%
\Delta }\left( 1+\dot{a}^{2}\right) +\frac{\Delta ^{3}}{a}\left( d-5\right) -%
\frac{6\Delta }{a}\left( a\ddot{a}+\frac{d-5}{2}\left( 1+\dot{a}^{2}\right)
\right) \right] \right\} ,$%
\end{tabular}%
\end{gather}%
where $\ell =\ddot{a}+f^{\prime }\left( a\right) /2$ and $\Delta =\sqrt{%
f\left( a\right) +\dot{a}^{2}}$ in which 
\begin{equation}
f\left( a\right) =\left. f_{-}\left( r\right) \right\vert _{r=a}.
\end{equation}%
We note that in our notation a 'dot' denotes derivative with respect to the
proper time $\tau $ and a 'prime' with respect to the argument of the
function. It can be checked by direct substitution from (28) and (29) that
the conservation equation 
\begin{equation}
\frac{d}{d\tau }\left( \sigma a^{\left( d-2\right) }\right) +p\frac{d}{d\tau 
}\left( a^{\left( d-2\right) }\right) =0.
\end{equation}%
holds true.

Once we know precisely the energy density and surface pressures, we can
study the energy conditions and the amount of exotic / normal matter that is
to support the above thin-shell wormhole. Let us start with the weak energy
condition (WEC) which implies for any timelike vector $V_{\mu }$ we must
have $T_{\mu \nu }V^{\mu }V^{\nu }\geq 0.$ Also by continuity, WEC implies
the null energy condition (NEC), which states that for any null vector $%
U_{\mu }$, $T_{\mu \nu }U^{\mu }U^{\nu }\geq 0$ \cite{2}. \ It is not
difficult to show that in an orthonormal basis these conditions read as%
\begin{equation}
\begin{array}{cc}
WEC: & \text{ }\rho \geq 0,\text{ }\rho +p_{i}\geq 0, \\ 
NEC & \rho +p_{i}\geq 0,%
\end{array}%
\end{equation}%
in which $i\in \left\{ 2,3,...,d-1\right\} .$ Here in the spherical
thin-shell wormholes, the radial pressure $p_{r}$ is zero and $\rho =\delta
\left( r-a\right) \sigma $ which imply WEC and NEC coincide as $\sigma \geq
0.$ Note that $\delta \left( r-a\right) $ stands for the Dirac
delta-function. By looking at $\sigma $ given in (28) one may conclude that
these conditions reduce to%
\begin{equation}
\frac{3}{2}a^{2}\leq \tilde{\alpha}\left( f\left( a\right) -2\dot{a}%
^{2}-3\right) .
\end{equation}%
For the static configuration with $\dot{a}=0,\ddot{a}=0$ and $a=a_{0}$ it is
not difficult to see that for $\tilde{\alpha}\geq 0$ the latter condition is
not satisfied. In other words, both WEC and NEC are violated. This is simply
from the fact that the metric function is asymptotically flat and $f\left(
a\right) <1$ for $a\geq r_{h}.$ Unlike $\tilde{\alpha}\geq 0,$ for the case
of $\tilde{\alpha}<0$ this condition in arbitrary dimensions is satisfied.
Direct consequence of these results can be seen in the total matter in
supporting the thin-shell wormhole. The standard integral definition of the
total matter is given by%
\begin{equation}
\Omega =\int \left( \rho +p_{r}\right) \sqrt{-g}d^{d-1}x
\end{equation}%
which gives 
\begin{equation}
\Omega =\frac{2\pi ^{\frac{d-1}{2}}a_{0}^{d-2}}{\Gamma \left( \frac{d-1}{2}%
\right) }\sigma _{0}
\end{equation}%
in which 
\begin{equation}
\sigma _{0}=-\frac{\sqrt{f\left( a_{0}\right) }\left( d-2\right) }{8\pi }%
\left[ \frac{2}{a_{0}}-\frac{4\tilde{\alpha}}{3a_{0}^{3}}\left( f\left(
a_{0}\right) -3\right) \right] .
\end{equation}%
It is obvious from $\Omega $ that similar to $\sigma _{0}$, in static
configuration the total matter which supports the thin-shell wormhole is
exotic if $\tilde{\alpha}\geq 0$ and normal if $\tilde{\alpha}<0.$ This
result is independent of dimensions and other parameters.

\section{Stability of the thin-shell wormholes for $d\geq 5$}

To study the stability of the thin-shell wormhole, constructed above, we
consider a radial perturbation of the radius of the throat $a.$ After the
linear perturbation we may consider a linear relation between the energy
density and radial pressure, namely \cite{12}%
\begin{equation}
p=p_{0}+\beta ^{2}\left( \sigma -\sigma _{0}\right) .
\end{equation}%
Here the constant $\sigma _{0}$ is given by (36) and $p_{0}$ reads as%
\begin{equation}
p_{0}=\frac{\sqrt{f\left( a_{0}\right) }}{8\pi }\left\{ \frac{2\left(
d-3\right) }{a_{0}}+\frac{f^{\prime }\left( a_{0}\right) }{f\left(
a_{0}\right) }-\frac{4\tilde{\alpha}}{a_{0}^{2}}\left[ \frac{f^{\prime
}\left( a_{0}\right) }{2}-\frac{f^{\prime }\left( a_{0}\right) }{2f\left(
a_{0}\right) }+\frac{f\left( a_{0}\right) \left( d-5\right) }{3a_{0}}-\frac{%
d-5}{a_{0}}\right] \right\} .
\end{equation}%
The constant parameter $\beta ^{2}$ for the wormhole supported by normal
matter is related to the speed of sound. By considering (37) in (31), one
finds 
\begin{equation}
\sigma \left( a\right) =\left( \frac{\sigma _{0}-p_{0}}{\beta ^{2}+1}\right)
\left( \frac{a_{0}}{a}\right) ^{\left( d-2\right) \left( \beta ^{2}+1\right)
}+\frac{\beta ^{2}\sigma _{0}-p_{0}}{\beta ^{2}+1}
\end{equation}%
in which $a_{0}$ is the radius of the throat in static equilibrium wormhole
and $\sigma _{0}(p_{0})$ is the static energy density (pressure) on the
thin-shell. By equating the latter expression and the one found by using
Einstein equation on the shell (28), we find the equation of motion of the
wormhole which reads 
\begin{equation}
\dot{a}^{2}+V\left( a\right) =0,
\end{equation}%
where%
\begin{equation}
V\left( a\right) =f\left( a\right) -\left( \left[ \sqrt{\mathbb{A}^{2}+%
\mathbb{B}^{3}}-\mathbb{A}\right] ^{1/3}-\frac{\mathbb{B}}{\left[ \sqrt{%
\mathbb{A}^{2}+\mathbb{B}^{3}}-\mathbb{A}\right] ^{1/3}}\right) ^{2}
\end{equation}%
and 
\begin{align}
\mathbb{A}& =\frac{3\pi a^{3}}{2\left( d-2\right) \tilde{\alpha}}\left[
\left( \frac{\sigma _{0}+p_{0}}{\beta ^{2}+1}\right) \left( \frac{a_{0}}{a}%
\right) ^{\left( d-2\right) \left( \beta ^{2}+1\right) }+\frac{\beta
^{2}\sigma _{0}-p_{0}}{\beta ^{2}+1}\right] , \\
\mathbb{B}& =\frac{a^{2}}{4\tilde{\alpha}}+\frac{1-f\left( a\right) }{2}.
\end{align}

Here $V\left( a\right) $ is called the potential of the wormhole's motion
and it helps us to figure out the regions of stability for the wormhole
under our linear perturbation. According to the standard method of stability
of thin-shell wormholes, we expand $V\left( a\right) $ as a series of $%
\left( a-a_{0}\right) .$ One can show that both $V\left( a_{0}\right) $ and $%
V^{\prime }\left( a_{0}\right) $ vanish and the first non-zero term in this
expansion is $\frac{1}{2}V^{\prime \prime }\left( a_{0}\right) \left(
a-a_{0}\right) ^{2}.$ Now, in a small neighborhood of the equilibrium point $%
a_{0}$ we have%
\begin{equation}
\dot{a}^{2}+\frac{1}{2}V^{\prime \prime }\left( a_{0}\right) \left(
a-a_{0}\right) ^{2}=0,
\end{equation}%
which implies that with $V^{\prime \prime }\left( a_{0}\right) >0,$ $a\left(
\tau \right) $ will oscillate about $a_{0}$ and make the wormhole stable. At
this point it will be in order also to clarify the status of parameter $%
\beta $ since ultimately the three-dimensional (i.e. $V^{\prime \prime
}\left( a_{0}\right) >0,$ $\beta ,$ $a_{0}$) stability plots will make use
of it. First of all although in principle $\beta <0$ is possible we shall
restrict ourselves only to the case $\beta >0.$ Unfortunately $\beta $ can
only be expressed implicitly as a function of $a_{0},$ through (37) and
expressions for $p,\sigma ,p_{0}$ and $\sigma _{0}$. It turns out that the
usual expression for stability, namely $V^{\prime \prime }\left(
a_{0}\right) >0$, can be plotted as a projection onto the plane formed by $%
\beta $ and $a_{0}$. This must not give the impression, however, that the
relation $\beta =\beta \left( a_{0}\right) $ is known explicitly.

\subsection{$d=5$}

Let us first eliminate $\alpha $ from the equations, by using the solution
given in (8). To do so we introduce new variables and parameters as%
\begin{eqnarray}
\tilde{a} &=&\frac{a}{\sqrt{\left\vert \alpha \right\vert }},\tilde{\tau}=%
\frac{\tau }{\sqrt{\left\vert \alpha \right\vert }},\tilde{Q}^{2}=\frac{Q^{2}%
}{\left\vert \alpha \right\vert },  \notag \\
\tilde{m} &=&\frac{2M_{ADM}}{3\left\vert \alpha \right\vert }+\frac{Q^{2}}{%
2\left\vert \alpha \right\vert }\ln \left\vert \alpha \right\vert .
\end{eqnarray}%
Upon these changes of variables, the other quantities change according to%
\begin{eqnarray}
f\left( a\right) &=&f\left( \tilde{a}\right) ,\sigma \left( a\right) =\frac{%
\sigma \left( \tilde{a}\right) }{\sqrt{\left\vert \alpha \right\vert }}%
,p\left( a\right) =\frac{p\left( \tilde{a}\right) }{\sqrt{\left\vert \alpha
\right\vert }},  \notag \\
\mathbb{A}\left( a\right) &=&\mathbb{A}\left( \tilde{a}\right) ,\mathbb{B}%
\left( a\right) =\mathbb{B}\left( \tilde{a}\right) ,V\left( a\right)
=V\left( \tilde{a}\right) .
\end{eqnarray}%
Finally the wormhole equation reads%
\begin{equation}
\left( \frac{d\tilde{a}}{d\tilde{\tau}}\right) ^{2}+\tilde{V}\left( \tilde{a}%
\right) =0.
\end{equation}%
Now, we consider two distinct cases, for $\alpha >0$ and $\alpha <0$,
separately.

\subsubsection{with $\protect\alpha >0$}

\bigskip In this section we consider $\alpha >0,$ such that the negative
branch of the EYM black hole solution reads%
\begin{equation}
f\left( \tilde{a}\right) =1+\frac{\tilde{a}^{2}}{4}\left( 1-\sqrt{1+\frac{16%
\tilde{m}}{\tilde{a}^{4}}+\frac{16\tilde{Q}^{2}\ln \tilde{a}}{\tilde{a}^{4}}}%
\right) ,
\end{equation}%
in which the condition 
\begin{equation}
\left. 1+\frac{16\tilde{m}}{\tilde{a}^{4}}+\frac{16\tilde{Q}^{2}\ln \tilde{a}%
}{\tilde{a}^{4}}\right\vert _{\tilde{a}=\tilde{a}_{0}}\geq 0,
\end{equation}%
and%
\begin{equation}
\left. \mathbb{A}^{2}+\mathbb{B}^{3}\right\vert _{\tilde{a}=\tilde{a}%
_{0}}\geq 0
\end{equation}%
must hold. The latter equation automatically is valid and the final relation
between the parameters reduces to (49). Based on this solution we find $%
\tilde{V}^{\prime \prime }\left( \tilde{a}_{0}\right) $ in terms of the
other parameters. Fig. 1 shows the stability regions and also $f\left( 
\tilde{a}\right) $ and $\sigma \left( \tilde{a}_{0}\right) .$

\subsubsection{with $\protect\alpha <0$}

\bigskip Next, we concentrate on the case $\alpha <0.$ With this choice
negative branch of the EYM black hole solution reads%
\begin{equation}
f\left( \tilde{a}\right) =1-\frac{\tilde{a}^{2}}{4}\left( 1-\sqrt{1-\frac{16%
\tilde{m}}{\tilde{a}^{4}}-\frac{16\tilde{Q}^{2}\ln \tilde{a}}{\tilde{a}^{4}}}%
\right) .
\end{equation}%
Based on this solution we study $\tilde{V}^{\prime \prime }\left( \tilde{a}%
_{0}\right) $ in terms of the other parameters. Fig.s 4 and 5 show the
stability regions and also $f\left( \tilde{a}\right) $ and $\sigma \left( 
\tilde{a}_{0}\right) .$

In this case also we have some constraint on the parameters in order to get $%
f\left( \tilde{a}_{0}\right) \geq 0,$ \ $\sigma \left( \tilde{a}_{0}\right)
\geq 0,$ and $\left. \mathbb{A}^{2}+\mathbb{B}^{3}\right\vert _{\tilde{a}=%
\tilde{a}_{0}}\geq 0.$ \ It is not difficult to see that all these
conditions reduce to 
\begin{equation}
0\leq \frac{1}{4}\sqrt{1-\frac{16\tilde{m}}{\tilde{a}_{0}^{4}}-\frac{16%
\tilde{Q}^{2}\ln \tilde{a}_{0}}{\tilde{a}_{0}^{4}}}\leq \frac{4}{\tilde{a}%
_{0}^{2}}-1,
\end{equation}%
and%
\begin{equation}
1-\frac{16\tilde{m}}{\tilde{a}_{0}^{4}}-\frac{16\tilde{Q}^{2}\ln \tilde{a}%
_{0}}{\tilde{a}_{0}^{4}}\geq 0.
\end{equation}%
After some manipulation, the parameters must satisfy the following constraint%
\begin{equation}
\tilde{a}_{0}^{4}\geq 16\left( \tilde{m}+\tilde{Q}^{2}\ln \tilde{a}%
_{0}^{2}\right)
\end{equation}%
where $0\leq \tilde{a}_{0}^{2}\leq 4.$ The stability region for this case is
given in Fig. 2.

\subsection{$d\geq 6$}

Here also we eliminate $\tilde{\alpha}$ from the equations. By introducing 
\begin{eqnarray}
\tilde{a} &=&\frac{a}{\sqrt{\left\vert \tilde{\alpha}\right\vert }},\tilde{%
\tau}=\frac{\tau }{\sqrt{\left\vert \tilde{\alpha}\right\vert }},\tilde{Q}%
^{2}=\frac{Q^{2}}{\left\vert \tilde{\alpha}\right\vert },  \notag \\
\tilde{m} &=&\frac{M_{ADM}}{\left\vert \tilde{\alpha}\right\vert ^{\frac{d-3%
}{2}}}.
\end{eqnarray}%
the other quantities become%
\begin{eqnarray}
f\left( a\right) &=&f\left( \tilde{a}\right) ,\sigma \left( a\right) =\frac{%
\sigma \left( \tilde{a}\right) }{\sqrt{\left\vert \alpha \right\vert }}%
,p\left( a\right) =\frac{p\left( \tilde{a}\right) }{\sqrt{\left\vert \alpha
\right\vert }},  \notag \\
\mathbb{A}\left( a\right) &=&\mathbb{A}\left( \tilde{a}\right) ,\mathbb{B}%
\left( a\right) =\mathbb{B}\left( \tilde{a}\right) ,V\left( a\right)
=V\left( \tilde{a}\right) ,
\end{eqnarray}%
and the wormhole equation is given by%
\begin{equation}
\left( \frac{d\tilde{a}}{d\tilde{\tau}}\right) ^{2}+\tilde{V}\left( \tilde{a}%
\right) =0.
\end{equation}

\subsubsection{with $\protect\alpha >0$}

\bigskip In this section we consider $\alpha >0,$ such that the negative
branch of the EYMGB black hole solution reads%
\begin{equation}
f_{-}\left( \tilde{a}\right) =1+\frac{\tilde{a}^{2}}{2}\left( 1-\sqrt{1+%
\frac{16\tilde{m}}{\tilde{a}^{d-1}\left( d-2\right) }+\frac{4\left(
d-3\right) \tilde{Q}^{2}}{\left( d-5\right) \tilde{a}^{4}}}\right)
\end{equation}%
Here we comment that constraints always restrict our free parameters. In the
case of $\alpha >0$ the first constraint is given by 
\begin{equation}
\left. \mathbb{A}^{2}+\mathbb{B}^{3}\right\vert _{\tilde{a}=\tilde{a}%
_{0}}\geq 0,
\end{equation}%
which upon substitution and manipulation automatically is satisfied for all
value of parameters. Based on this solution we find $\tilde{V}^{\prime
\prime }\left( \tilde{a}_{0}\right) $ in terms of the other parameters.
Fig.s 3-5 shows the stability regions and also $f\left( \tilde{a}\right) $
and $\sigma \left( \tilde{a}_{0}\right) $ for dimensions $d=6,7$ and $8.$

\subsubsection{with $\protect\alpha <0$}

\bigskip Next, we concentrate on the case $\alpha <0.$ With this choice
negative branch of the EYMGB black hole solution reads%
\begin{equation}
f_{-}\left( \tilde{a}\right) =1-\frac{\tilde{a}^{2}}{2}\left( 1-\sqrt{1-%
\frac{16\tilde{m}}{\tilde{a}^{d-1}\left( d-2\right) }-\frac{4\left(
d-3\right) \tilde{Q}^{2}}{\left( d-5\right) \tilde{a}^{4}}}\right) .
\end{equation}%
Based on this solution we study $\tilde{V}^{\prime \prime }\left( \tilde{a}%
_{0}\right) $ in terms of the other parameters. Fig. 6 show the stability
regions and also $f\left( \tilde{a}\right) $ and $\sigma \left( \tilde{a}%
_{0}\right) .$ In order to set $f\left( \tilde{a}_{0}\right) \geq 0,$ \ $%
\sigma \left( \tilde{a}_{0}\right) \geq 0,$ and $\left. \mathbb{A}^{2}+%
\mathbb{B}^{3}\right\vert _{\tilde{a}=\tilde{a}_{0}}\geq 0$ \ it is enough
to satisfy 
\begin{equation}
0<\sqrt{1-\frac{16\tilde{m}}{\tilde{a}_{0}^{d-1}\left( d-2\right) }-\frac{%
4\left( d-3\right) \tilde{Q}^{2}}{\left( d-5\right) \tilde{a}_{0}^{4}}}<%
\frac{2}{\tilde{a}_{0}^{2}}-1,
\end{equation}%
and%
\begin{equation}
1-\frac{16\tilde{m}}{\tilde{a}^{d-1}\left( d-2\right) }-\frac{4\left(
d-3\right) \tilde{Q}^{2}}{\left( d-5\right) \tilde{a}^{4}}>0,
\end{equation}%
where $0<\tilde{a}_{0}^{2}<2.$

\section{CONCLUSION}

We have investigated the possibility of thin-shell wormholes in EYMGB theory
in higher ($d\geq 5$) dimensions with particular emphasis on stability
against spherical, linear perturbations and normal (i.e. non-exotic) matter.
For this purpose we made use of the previously obtained solutions that are
valid in all dimensions. The case $d=5$ is considered separately from the
cases $d>5$ because the solution involves a logarithmic term apart from the
power-law dependence. For $d=5$ we observe (Fig. 2) the formation of a
narrow band of positive energy region that attains a stable wormhole only
for $\alpha <0.$ On the contrary, for $\alpha >0$ although a large region of
stability (i.e. $V^{\prime \prime }\left( a_{0}\right) >0$) forms, the
energy turns out to be exotic. For $d>5$ also, we have more or less a
similar picture. That is, whenever the GB parameter $\alpha >0,$ negative
energy shows itself versus the stability requirements. We have analyzed the
cases $d=6,7$ and $8$ as examples. Our technique is powerful enough to apply
in any higher dimensions, however, for technical reasons we had to be
satisfied with these selected dimensions. We must admit also that for
non-spherical perturbations a similar analysis remains to be seen. In our
study we were able also to observe a stability region which employs $0<\beta
<1,$ which can be interpreted as a case corresponding to less than the speed
of light. In conclusion, we state that formation of stable, positive energy
thin-shell wormholes in EYMGB are possible only with a GB parameter $\alpha
<0.$ Without the GB term whatever source is available the situation is
always worse. The indispensable character of the GB parameter toward useful
wormhole constructions invites naturally the Lovelock hierarchy \cite{5} for
which GB term constitutes the first member. 

\bigskip

\textbf{Figure\textbf{\ caption}s:}

Figure 1: Region of stability (i.e. $V^{\prime \prime }\left( a_{0}\right)
>0 $) for the thin-shell in $d=5$ and for $\alpha >0$. The $f\left( r\right) 
$ and $\sigma _{0}$ plots are also given. It can easily be seen that the
energy density $\sigma _{0}$ is negative which implies exotic matter.

Figure 2: For $d=5$ and $\alpha <0$ case with the chosen parameters $f\left(
r\right) $ has no zero but $\sigma _{0}$ has a small band of positivity with
the presence of normal matter. We note also that $\beta <1$ in a small band.

Figure 3: For $d=7$ and $\alpha >0$ the stability region is plotted which is
seen to have exotic matter alone.

Figure 4: For $d=6$ and $\alpha >0$ also a region of stability is available
but with $\sigma _{0}<0$. Note that $d=6$ is special, since from Eq. (5) in
the text we have $\kappa =0$ and the energy-momentum takes a simple form.

Figure 5: For $d=8$ with $\alpha >0$ exotic matter is seen to be
indispensable.

Figure 6: For $d=6$ with $\alpha <0$ there are two disjoint regions of
stability for the thin-shell and in contrast to the $\alpha >0$ case in Fig.
4, we have $\sigma _{0}>0.$ We notice in this case also that $\beta <1$ is
possible.

\end{document}